\documentclass[11pt,a4paper]{article}
\pdfoutput=1

\usepackage{jheppub}
\usepackage{chessfss}
\usepackage{dsfont}

\newcommand{\beqa}{\begin{eqnarray}}
\newcommand{\eeqa}{\end{eqnarray}}

\def\Tr{\rm Tr}

\newcommand{\beq}{\begin{equation}}
\newcommand{\eeq}{\end{equation}}
\newcommand{\bea}{\begin{eqnarray}}
\newcommand{\eea}{\end{eqnarray}}

\newcommand{\CN}{{\mathcal N}}

\newcommand{\be}{\begin{equation}}
\newcommand{\ee}{\end{equation}}

\def\+{{+\!\!\!+}}

\def\0{\nonumber}

\def\Tr{{\rm Tr}}

\def\1{{\bf 1}}

\setboardfontsize{8}

\title{Four dimensional superconformal theories from $M5$ branes.}

\author{Simone Giacomelli}

\affiliation{Universit\'e Libre de Bruxelles
and International Solvay Institutes\\
ULB-Campus Plaine CP231
1050 Brussels, Belgium.}

\emailAdd{simone.giacomelli@ulb.ac.be}

\abstract{
   We study N=1 superconformal theories in four dimensions obtained wrapping M5 branes on a Riemann surface. We propose 
a method to determine from the spectral curve the scaling dimension of chiral operators in the SCFT. Whenever the R-symmetry 
has to be determined via a-maximization, our procedure allows us to determine the charge of chiral operators under the ``trial'' 
R-symmetry. Our proposal reduces to the correct prescription in the special case of N=2 theories of class $\mathcal{S}$. We perform several 
consistency checks and apply our method to study some new SCFT's such as N=1 deformations of Argyres-Douglas theories.
}

\keywords{}

\begin{document}
\setcounter{tocdepth}{2}
\maketitle
\section{Introduction}

 In the past two decades it has become clear that the brane engineering of supersymmetric gauge theories provides a convenient 
setup in which much of the information about the protected sector of the field theory has a simple geometric interpretation. This 
often helps in simplifying the study of the field theory and makes more transparent the various dualities linking different-looking 
field theories. 

Among the various examples studied so far we would like to mention the M5 brane description of $\CN=2$ gauge theories in four dimensions 
proposed by Witten in \cite{witten2}. In this case the geometry of the M5 brane encodes all the information about the low energy dynamics of the gauge 
theory, providing a framework in which the Seiberg-Witten curve of the theory can be easily determined. This simplifies considerably the 
field theory analysis pioneered in \cite{SW1,SW2}. More recently this result was generalized by Gaiotto in \cite{gaiotto}, where it was recognized 
that a large class of $\CN=2$ superconformal theories can be constructed by compactifying on a Riemann surface the $\CN=(2,0)$ 
theory of type $A_{N-1}$. 

This construction provides a simple geometric framework which allows us to study in a large class of theories 
Argyres-Seiberg like dualities \cite{AS}. Another lesson we can learn from \cite{gaiotto} is that starting from six dimensions we 
can construct many new $\CN=2$ superconformal theories without any obvious lagrangian description. In studying these non conventional 
theories, one of the most powerful tools at our disposal is the Seiberg-Witten curve which allows us to extract information 
about the chiral ring of the theory. In particular, the structure of the curve combined with the observation that the Seiberg-Witten 
differential has dimension one for every $\CN=2$ SCFT (just because the periods of the Seiberg-Witten differential give 
the mass of BPS states), allows us to determine the scaling dimension of chiral operators in the CFT \cite{AD,ADSW}. 

Recently, the analysis of \cite{gaiotto} has been extended to include a large class of superconformal theories with four supercharges in 
\cite{dann1,kazuya}. The construction of these $\CN=1$ theories involves compactifying a stack of $M5$ branes on a Riemann 
surface as in \cite{gaiotto}. The difference with respect to the $\CN=2$ case is that the M-theory background is 
$\mathbb{R}^5\times CY_3$ instead of $\mathbb{R}^7\times CY_2$. In both cases the Riemann surface is a holomorphic cycle in the 
Calabi-Yau manifold (see also \cite{BTW}-\cite{Index} for earlier work on this topic). 

As explained in \cite{dankazuya} (see also 
\cite{noi}) a spectral curve encoding the properties of the chiral ring, analogous to the Seiberg-Witten curve for $\CN=2$ 
theories, can be written down for $\CN=1$ theories obtained compactifying $M5$ branes on a surface.  
This curve can be identified with the one first introduced by Intriligator and Seiberg in \cite{IS1} (see also \cite{IS2,IS3}), which determines the holomorphic 
gauge coupling of low energy massless $U(1)$ fields in $\CN=1$ Coulomb phase. In \cite{dankazuya} it was checked that when one 
restricts to the models considered in \cite{IS1,IS2,IS3}, the $\CN=1$ spectral curve reduces to the Intriligator-Seiberg curve 
found in these papers.

It is then natural to ask whether this curve allows us to fix the scaling dimensions of chiral operators as in the $\CN=2$ case. This 
raises immediately a few questions: frequently in $\CN=1$ theories the scaling dimensions are irrational. This will generically 
be the case whenever the exact R-symmetry has to be determined via a-maximization \cite{amax}. How can we derive such a result 
from a curve which is defined as the zero locus of polynomial equations? The second question is what plays the role 
of the Seiberg-Witten differential in the $\CN=1$ case. The purpose of this note is to address these questions. 

We find that 
whenever the $U(1)$ R-symmetry is not uniquely fixed and we have to apply a-maximization, the curve allows us to fix the ``trial'' 
scaling dimensions (before maximizing the trial a central charge). The final result will then be found by evaluating the trial a 
central charge and finding its maximum. As is well known, it might happen that some operators apparently violate the unitarity bound ($D(O)>1$) as a result 
of this procedure. Often this signals the fact that these operators decouple and become free \cite{Seiberg}. This can be included in 
the a-maximation following the prescription of \cite{kutasov}. 

The second question can be answered exploiting the observation made in \cite{witten} 
(see also \cite{kazuya}) that the integral of the holomorphic top form of $CY_3$ gives the effective superpotential. We should then 
require that the holomorphic three-form has scaling dimension three. We will check in particular that this prescription allows us to 
recover the correct constraint for $\CN=2$ theories of class $\mathcal{S}$: the SW differential has dimension one. 

In section 2 we provide a short review of class $\mathcal{S}$ theories, focussing on the results that we will need in 
later sections. In section 3 we introduce the spectral curve for $\CN=1$ theories and discuss the role of the holomorphic three-form. 
We then explain how to determine the trial a central charge. This latter computation is closely related to the analysis performed 
in \cite{BBBW2,TMWG}, which we apply. In section 4 we perform some consistency checks of our proposal and discuss 
lagrangian theories in this class. We analyze in detail two special cases which have not been discussed so far. In 
section 5, using the results derived in \cite{noi}, we apply our method to $\CN=2$ theories deformed by an $\CN=1$ preserving 
superpotential. In particular we study $\CN=1$ deformations of Argyres-Douglas theories of type $A_{N-1}$. We conclude with some 
final comments in section 6.
\vspace{10pt}\\
{\bf Note added:} While completing this work, \cite{BK} appeared on the arXiv. The two models we discuss in section 4.3.2 are special 
cases of the theories studied in that paper.

\section{Four dimensional SCFTs via M5 branes}

In this section we briefly review the main features of theories of class $\mathcal{S}$ (both $\CN=2$ and $\CN=1$). We refer to 
e.g. \cite{revt} and \cite{kazuya} for more details. The readers familiar with these topics can skip this section.

\subsection{$\CN=2$ theories from M5 branes}

\subsubsection{Topological twist and Seiberg-Witten curve}

As is well known, the 6d $\CN=(2,0)$ theories can be compactified (with a suitable topological twist) on 
a Riemann surface $\mathcal{C}$ with punctures in such a way that the resulting 4d theory preserves 8 supercharges. The four 
dimensional theories one gets in this way are usually referred to as class $\mathcal{S}$ theories.
In the $A_{N-1}$ case studied in \cite{gaiotto} this procedure corresponds to cosidering M-theory on the background 
$\mathbb{R}^4\times\mathbb{R}^3\times CY_2$, where $CY_2$ can be thought of as the total space of the cotangent bundle on $\mathcal{C}$ 
and introducing a stack of N M5 branes wrapping $\mathbb{R}^4\times\mathcal{C}$, and sitting at the origin\footnote{More 
precisely, this is strictly true on the Coulomb branch of the theory. The motion in the transverse $\mathbb{R}^3$ describes the 
motion onto the Higgs branch of the theory. See \cite{DK} for a recent discussion on this point.} of $\mathbb{R}^3$. The 
invariance of the system under rotations in $\mathbb{R}^3$ is just the geometric counterpart of the $SU(2)_R$ symmetry of the 
theory. 

We can choose local coordinates z and x parametrizing $\mathcal{C}$ and the fiber of the cotangent bundle respectively, such that the 
holomorphic two-form of $CY_2$ has the form \cite{GMN} $$\Omega_2=dx\wedge dz=d(xdz)=d\lambda,$$ where $\lambda=xdz$ is the Seiberg-Witten 
differential. At low energy the N branes recombine into a single M5 brane wrapping $\mathbb{R}^4\times\Sigma$, where 
$\Sigma$ is an N-sheeted covering of $\mathcal{C}$ described by the equation 
$$\lambda^N=\sum_{k=2}^{N}\lambda^{N-k}\phi_k(z).$$ 
This is nothing else than the Seiberg-Witten curve encoding the low-energy effective action of the theory \cite{SW1, SW2} and $\phi_k$ 
are meromorphic k-differentials with poles at the punctures. There is a large variety of punctures, which can be characterized in terms of 
the degree of the poles of the k-differentials. We will mainly consider regular punctures in the following, which means 
$$\text{deg}\phi_k\leq k\; \forall k.$$ They can be nicely classified in terms of Young diagrams with N boxes, or equivalently by a partition 
of N.

\subsubsection{Punctures and flavor symmetries.}

The punctures we have introduced above describe the codimension-two defects of the 6d theory and their effect in the present context is to encode the 
flavor symmetry of the theory: each puncture is associated (see e.g. \cite{yujidist}) to an embedding $\rho$ of $SU(2)$ in $SU(N)$ such that $\rho(\sigma^+)$ 
is nilpotent with Jordan blocks of size $n_i$ (where $n_i$ indicates the height of the i-th column of the Young diagram). The 
commutant of this $SU(2)$ subgroup in $SU(N)$ gives the flavor symmetry associated with the puncture.

Two distinguished types of punctures are the full (or maximal) and simple (or minimal) punctures. The first is described by a 
Young diagram with a single row of length N. In this case the embedding of $SU(2)$ is trivial ($\rho(\sigma^+)=0$ since all 
Jordan blocks have dimension one) and the associated flavor symmetry is $SU(N)$. The second is associated with a Young diagram with 
a column of height $N-1$ and one of height one. In this case $\rho(\sigma^+)$ has a Jordan block of dimension $N-1$ and the 
corresponding flavor symmetry is just $U(1)$. More in general the flavor symmetry is $S(\prod_iU(r_i))$, where $r_i$ is the number 
of columns of height i. Under the above mentioned embedding, the fundamental representation of $SU(N)$ decomposes into irreducible 
representations of $SU(2)$ as $\mathbf{N}\rightarrow 
\sum_{i=1}^l\mathbf{n}_i$, where $\mathbf{n}_i$ is again the height of the i-th column. We can easily derive from this formula the decomposition 
of the adjoint of $SU(N)$ \cite{TMWG}:
\be\label{decom} \mathbf{adj.}=\bigoplus_{i=1}^{l}\bigoplus_{s=1}^{n_i-1}V_{s}\oplus(l-1)V_0\oplus2\left[ 
\bigoplus_{i<j}\bigoplus_{k=1}^{n_j}V_{\frac{n_i+n_j-2k}{2}}\right]\equiv\bigoplus_s R_s V_s,\ee 
where $V_{s}$ is the spin $s$ representation of $SU(2)$ and $R_s$ denote the flavor symmetry representations. 

The chiral ring of a class $\mathcal{S}$ theories includes for every puncture a multiplet $\mu$ transforming in the adjoint of the 
corresponding flavor symmetry (the moment map associated with the global symmetry). Starting from a 
theory described by a surface with full punctures only we can obtain any other class $\mathcal{S}$ model with the same number of punctures 
by giving the suitable nilpotent vev to the $\mu_i$ fields: 
$$\langle\mu_i\rangle=\rho_i(\sigma^+),$$ where $\rho_i$ indicate the $SU(2)$ embeddings described before. So, if we want to get 
e.g. a simple puncture, $\langle\mu_i\rangle$ should have a Jordan block of dimension $N-1$. This operation is 
usually called closure of the puncture. The process produces for every puncture $N_p$ free half-hypers, where $N_p$ is the dimension of the 
orbit of $\rho_p(\sigma^+)$. From (\ref{decom}) we find 
\be N_p=\sum_{i=1}^{l}\sum_{s=1}^{n_i-1}2s+\sum_{i<j}\sum_{k=1}^{n_j}2(n_i+n_j-2k).\ee 

\subsubsection{Central charges and anomalies} 

Given a superconformal theory with a global symmetry $G$, the flavor central charge $K$ of $G$ is defined as \cite{anselmi, anselmi2} 
\be\label{ciao} K_G\delta^{ab}=-3\Tr RT^aT^b,\ee where $R$ is the R-symmetry of the theory and $T^a$, $T^b$ are the generators 
of $G$. For $\CN=2$ SCFT's the usual definition is
\be\label{ciao1}K_G\delta^{ab}=-2\Tr R_{\CN=2}T^aT^b.\ee
We adopt the same conventions as in \cite{TMWG}: the quadratic Casimirs of $SU(N)$ are $\frac{1}{2}$ and $N$ for the 
fundamental and adjoint representations respectively. In class $\mathcal{S}$ theories the central charge of a flavor symmetry associated with 
a puncture is given by the formula \cite{yujidist} 
$$K\delta^{ab}=2\sum_s \Tr_{R_s}T^aT^b,$$ where $R_s$ are the representations appearing in (\ref{decom}). For the 
$SU(r_i)$ subgroups introduced in the previous section this formula reduces to 
\be\label{globcc}K_{SU(r_i)}=2\sum_{j\leq i}l_j,\ee where $l_j$ is the length of the j-th row of the Young diagram (notice that 
$r_i=l_i-l_{i+1}$). 

In the following we will need to evaluate the a and c central charges, which for a superconformal theory can be expressed in 
terms of the anomalies of the R current \cite{anselmi2}: 
\be\label{ac}a=\frac{3}{32}(3\Tr R^3-\Tr R);\quad c=\frac{1}{32}(9\Tr R^3-5\Tr R).\ee 
For $\CN=2$ theories it is convenient to introduce the parameters $n_v$ and $n_h$ defined as follows \cite{GM}: 
$$a=\frac{5n_v+n_h}{24};\quad c=\frac{2n_v+n_h}{12}.$$ For free theories these coincide with the number of vectormultiplets and 
hypermultiplets respectively. Notice that the positivity of these two parameters is equivalent to the unitarity constraint found in 
\cite{ST} for $\CN=2$ SCFTs: $$\frac{1}{2}\leq\frac{a}{c}\leq\frac{5}{4}.$$

For class $\mathcal{S}$ theories these can be read off from the data of the Riemann surface \cite{CD}: they receive a global 
contribution which can be computed using the anomaly polynomial of the 6d theory (we will discuss this in the next section) and 
a local contribution from each puncture, which is given by the following formula for the 6d theory of type $A_{N-1}$: 
\be\label{nseff}n_v=\sum_{k=2}^{N}(2k-1)p_k;\quad n_h=n_v+\frac{1}{2}(\sum_jl_j^2-N),\ee 
where $l_j$ is again the length of the j-th row of the Young diagram and $p_k$ is the degree of the pole of the k-th meromorphic 
differential.

\subsection{$\CN=1$ theories from M5 branes}

Let us consider now M-theory on the background $\mathbb{R}^4\times X\times\mathbb{R}$, where X is a Calabi-Yau threefold. 
A stack of N M5 branes wrapping $\mathbb{R}^4\times\mathcal{C}$, where $\mathcal{C}$ is a holomorphic two-cycle in X, describes an $\CN=1$ theory on 
$\mathbb{R}^4$. Although more general choices are possible, we will restrict to the case $$X=\mathcal{L}_1\oplus\mathcal{L}_2,$$ 
where $\mathcal{L}_1$ and $\mathcal{L}_2$ are holomorphic line bundles on $\mathcal{C}$ of degree $p$ and $q$. Indeed the Calabi-Yau 
condition imposes the constraint $p+q=2g-2$, where $g$ is the genus of the Riemann surface. 

A large class of $\CN=1$ SCFTs of this kind, associated with Riemann surfaces with punctures, have been constructed in \cite{dann1}.
The basic building blocks are three-punctured spheres of two different kinds (we will call them black and red). These can be 
connected together to form higher genus Riemann surfaces. As in the $\CN=2$ case, connecting together two spheres corresponds to 
gauging the diagonal subgroup of the corresponding flavor symmetries. If the two spheres are of the same kind the vectormultiplet 
will be $\CN=2$, otherwise it will be $\CN=1$. As noticed in \cite{dann1} (see also \cite{TMWG}), for every $\CN=1$ gauging we have 
the superpotential term $$\Tr\mu_1\mu_2,$$ where $\mu_1$ and $\mu_2$ are the moment maps associated with the flavor symmetries. When the  
vectormultiplet is $\CN=2$ we have instead the standard superpotential term $$\Tr\Phi(\mu_1-\mu_2),$$ where $\Phi$ is the chiral 
multiplet in the adjoint. In both cases the complex moduli of the Riemann surface are identified with 
exactly marginal parameters of the field theory. In the $\CN=2$ case these correspond to gauge couplings whereas in the $\CN=1$ 
case these are the couplings associated with the above superpotential terms.

\begin{figure}
\centering{\includegraphics[width=.5\textwidth]{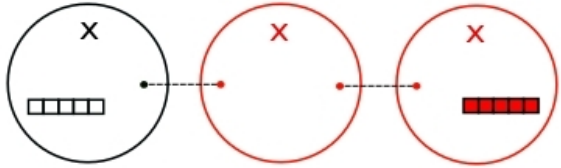}} 
\caption{\label{lin}\emph{We have a linear quiver with two $SU(N)$ gauge groups (in this case $N=5$). One vectormultiplet is 
$\CN=1$, since we connect spheres of different kind, whereas the other is $\CN=2$. We indicated two full punctures with the corresponding 
Young diagram. The cross denotes a simple puncture and the dashed lines denote the tubes connecting the various three-punctured 
spheres. The dots indicate full punctures whose $SU(N)$ flavor symmetry is gauged. All the three-punctured spheres in the figure describe 
bifundamentals of $SU(N)\times SU(N)$.}}
\end{figure}

There is also a large family of punctures labelled by ``decorated'' Young diagrams, 
which are closely related to a subset of the 1/4 BPS boundary conditions studied in \cite{masa,masa1}. One distinguished subclass 
is the set of punctures which preserve 8 supercharges. Also these come in two groups and we will refer to them as black and red. 
These two families of punctures preserve different subalgebras of the $\CN=(2,0)$ superalgebra so, when both kinds of punctures are 
present, the corresponding four dimensional theory will inherit only 4 supercharges. In this note we will restrict ourselves to 
theories with these $\CN=2$ preserving punctures only (see figure \ref{lin} for an example).

A geometric description of these models has been proposed in \cite{kazuya}: we decompose the punctured Riemann surface into spheres 
with three holes (pair of pants) and caps with a puncture. For each of these building blocks we take the canonical and the 
trivial line bundles. For a ``black'' building block we identify the canonical bundle with $\mathcal{L}_1$ and the trivial bundle with $\mathcal{L}_2$. 
For a ``red'' building block we make the opposite identification. Each black (red) cap contributes +1 to the integral of the 
first Chern class of $\mathcal{L}_1$ ($\mathcal{L}_2$) and each sphere with three holes contributes -1. When we connect two building blocks of the 
same kind we glue the corresponding canonical bundles (and analogously we glue together the corresponding trivial bundles). If the 
building blocks are instead of different type we glue the canonical bundle of the first to the trivial bundle of the second. We 
easily see that if all building blocks are of the same kind (let's say black), $\mathcal{L}_1$ gets identified with the canonical bundle of the 
Riemann surface $\mathcal{C}$ and the threefold is of the form $T^*(\mathcal{C})\times\mathbb{C}$. This special case corresponds to $\CN=2$ 
theories of class $\mathcal{S}$. 

We thus find that our $\CN=1$ theories are labelled by the number of black and red spheres ($n_B$ and $n_R$) and by the number of black 
and red caps ($p_B$ and $p_R$). From these data we can determine the degree of the two line bundles: 
\be\label{degree}\text{deg}\; \mathcal{L}_1 = p = n_{B}-p_B;\quad \text{deg}\; \mathcal{L}_2 = q = n_{R}-p_{R}.\ee
All the theories for which these four numbers are the same are believed to be dual to each other (see \cite{dann1},\cite{TMWG} and 
also \cite{Song}).

\section{Spectral curve and scaling dimensions}

\subsection{$\CN=1$ curves and holomorphic three-form}

For the class of theories described above we can write a spectral curve as in the $\CN=2$ case. It can be written in the form\footnote{
This is the spectral curve associated with a pair of commuting Hitchin fields. The third equation is introduced to make the curve an 
N-sheeted covering of the UV curve, which is a natural requirement since the curve describes a stack of $N$ M5 branes wrapping 
the UV curve. As noticed in \cite{dankazuya} this equation follows from the commutativity constraint under the assumption that the 
roots of the second equation are distinct at a generic point on the UV curve. When $N=2$ this assumption is not necessary.}  
\begin{equation}\label{spec}\begin{cases} s_1^N=\sum_{k=2}^{N}s_1^{N-k}\phi_{1k}(z)\\
 s_2^N=\sum_{k=2}^{N}s_2^{N-k}\phi_{2k}(z)\\
 s_1=\varphi_{N-1}(z)s_2^{N-1}+\dots+\varphi_{0}(z)
\end{cases} \end{equation}
where $s_1$ is a section of $\mathcal{L}_1$ and $s_2$ is a section of $\mathcal{L}_2$. $\phi_{ik}$ are meromorphic 
sections of $\mathcal{L}_i^{\otimes k}$ with poles at the punctures. This parametrization of the curve was found in \cite{noi} for $\CN=2$ 
theories associated with a linear quiver of three-punctured spheres deformed by a polynomial superpotential for the chiral multiplets 
in the adjoint and extended to the class of models described in the previous section in \cite{dankazuya}. The idea is that 
$\phi_{1k}$ have poles of degree $p_k$ (the same as k-differentials in $\CN=2$ theories) at black punctures and do not diverge at 
red punctures (same story for $\phi_{2k}$ with the roles of black and red punctures interchanged).

In the study of $\CN=2$ theories, assigning the curve is not enough to identify the theory: we must also specify the Seiberg-Witten 
differential. To illustrate this point, consider e.g. the curve $y^2=x^N+\dots+u_N$. Depending on whether the SW differential is 
$ydx$ or $(y/x)dx$, this describes the Argyres-Douglas theory of type $A_{N-1}$ or $D_N$ (see e.g. \cite{EIHY}).

Analogously, in the present case we need to specify both the curve and the holomorphic three-form of X. Consider for instance $SU(N)$ SYM theory. 
In this case the curve $\mathcal{C}$ is a sphere with two irregular punctures and the two line bundles have degree -1. The curve can be written 
as follows \cite{noi} (setting to one the dynamical scale)
\begin{equation}\begin{cases} x_1^N=1/z\\
 x_2^N=z\\
 x_1x_2=1 \end{cases}\end{equation} 
This is the curve already found by Witten \cite{witten}, who also observed that the holomorphic three-form has the form 
$$\Omega=dx_1\wedge dx_2\wedge \frac{dz}{z}.$$ 
Following the recipe proposed in \cite{dankazuya} we find instead the curve 
\begin{equation}\begin{cases} x_1^N=1/z^{N+1}\\
 x_2^N=z\\
 x_1x_2=1/z \end{cases}\end{equation}
The two curves look different but indeed one can go from the first to the second with a redefinition of $x_1$. Equivalently, we can 
say that the difference is in the normalization of the holomorphic three-form, which in the latter case is simply
$$\Omega=dx_1\wedge dx_2\wedge dz.$$

In a neighbourhood of a puncture (say black), the system is locally $\CN=2$ and we can choose local coordinates $z$, $x_1$, $x_2$ 
on $\mathcal{C}$ and the fibers of $\mathcal{L}_1$ and $\mathcal{L}_2$ such that $x_2$ does not diverge at the punture, the 
singularity of $x_1$ is the same as in \cite{dankazuya} and
$$\Omega=d\lambda\wedge dx_2=dx_1\wedge dx_2\wedge dz.$$ As emphasized in \cite{gaiotto}, for $\CN=2$ theories of class $\mathcal{S}$ there is a 
``canonical'' normalization for the SW differential, whereas in this case we have a canonical normalization for the holomorphic three-form.

As noticed by Witten in \cite{witten}, the effective superpotential can be computed integrating the holomorphic 
three-form $\Omega$ on a three cycle $B$ whose boundary is the union of $\Sigma$ (the spectral curve) and $\Sigma_0$ (a two cycle in the same homology class of 
$\Sigma$) $$\mathcal{W}=\int_B\Omega.$$ In $\CN=1$ theories that flow in the infrared to a 
superconformal fixed point, we expect the effective superpotential $\mathcal{W}$ to be exactly marginal, or equivalently 
that the three form $\Omega$ has scaling dimension three\footnote{Equivalently, we can demand that the 
superspace measure $d^2\theta$ and $\Omega^{-1}$ have the same R-charge, as in Calabi-Yau compactifications (see \cite{witten}). This leads to 
$R(\Omega)=2$ for superconformal theories.} \be\label{scaledif}[\Omega]=3.\ee This is our proposal for the $\CN=1$ counterpart of the constraint on the Seiberg-Witten 
differential for $\CN=2$ theories\footnote{As a side remark, we would like to mention the fact that a different formula for the 
effective superpotential was proposed in \cite{kazuya}. The rest of our analysis is unaltered if we adopt this formula as a starting 
point. As will become clear in later sections, this is simply a consequence of the ``canonical'' normalization for $\Omega$ introduced above.}.

\subsection{$\CN=1$ theories from Riemann surfaces with regular punctures}

All the theories associated with a surface with regular punctures only are believed to be superconformal\footnote{This is always 
true when the genus of $\mathcal{C}$ is one or higher. In the case of the sphere the situation is more subtle and the theory 
admits the standard interpretation we have discussed when we have at least three punctures satisfying a certain relation. We will 
discuss further this point for $\CN=2$ theories in section 4.}. In this class of models the coordinate 
$z$ parametrizing the surface has zero R-charge, or equivalently zero scaling dimension. The constraint on the holomorphic 
three-form then tells us that $[x_1]+[x_2]=3$, or equivalently that $R(x_1)+R(x_2)=2$. Notice that the curve (\ref{spec}) does not 
imply any further constraint on the scaling dimensions of $x_1$ and $x_2$\footnote{We are free to assign scaling dimensions to all the 
parameters appearing in the curve in such a way that the equations (\ref{spec}) are homogeneous.}. We are then lead to the equation 
\be\label{dimension}R(x_1)=1+\epsilon; \quad R(x_2)=1-\epsilon,\ee where the parameter $\epsilon$ is undetermined. 

We can now provide the first consistency check: let's consider a red puncture located at $z=z_r$, where $z$ is a local coordinate 
on the Riemann surface. In \cite{kazuya} it was argued that the limit for $z\rightarrow z_r$ of the spectral equation for $x_1$ (the first equation in (\ref{spec})) 
coincides with the characteristic polynomial of the corresponding moment map $\mu_{r}$\footnote{More precisely, this is true when the 
puncture and the sphere are of the same kind. This will be enough for the present argument. In the other case 
we have a different relation which will be discussed later.}. We thus conclude that $x_1$ and 
$\mu_{r}$ (for all values of $j$) have the same R-charge. An analogous relation holds for $x_2$ and $\mu_{b}$ at black 
punctures. With this assignment of R charges the superpotential terms $\Tr\mu_1\mu_2$ and $\Tr\Phi(\mu_1-\mu_2)$ which arise in this 
class of theories are automatically exactly marginal for $\CN=1$ and $\CN=2$ gauge groups respectively. In the $\CN=1$ case this statement 
directly follows from the above discussion. In the $\CN=2$ case we also need the fact that locally one of the two coordinates 
can be seen as parametrizing the cotangent bundle on the surface and this has the same scaling dimension as $\Phi$.

Notice that the R-charges in (\ref{dimension}) coincide with the charges of the two coordinates under the $U(1)$ group 
$$R_{\epsilon}=R_0+\epsilon F,$$ where 
$R_0$ and $F$ are respectively the diagonal and antidiagonal combinations of the two $U(1)$ groups which act as phase rotations 
on the fibers of the two line bundles.
The idea now is to fix the parameter $\epsilon$ exploiting a-maximization \cite{amax}. The problem is thus reduced to computing the trial 
central charge $a(\epsilon)$. As we already mentioned in section 2,
the a central charge is given by the sum of a global contribution, which can be determined using the anomaly 
polynomial for the $\CN=(2,0)$ theory, and a local contribution from each puncture. 
The global contribution has been determined in \cite{BBBW2}. Introducing the parameter $$z=\frac{p-q}{2g-2},$$ where p and q 
are the degrees of the two line bundles given by (\ref{degree}), we find 
$$\Tr R_{\epsilon}=(g-1)r_G(1+z\epsilon);\quad \Tr R^3_{\epsilon}=(g-1)[(r_G+d_Gh_G)(1+z\epsilon^3)-d_Gh_G(\epsilon^2+z\epsilon)],$$
where $r_G$, $h_G$ and $d_G$ are respectively the rank, Coxeter number and dimension of the simply-laced ADE groups. Notice that 
the presence of punctures affects this computation, since the value of the $z$ parameter depends on them. 
Using then (\ref{ac}) we find 
\be\label{apoly}a_{global}(\epsilon)=\frac{3}{32}(g-1)\left[3(r_G+d_Gh_G)z\epsilon^3-3d_Gh_G\epsilon^2-(r_G+3d_Gh_G)z\epsilon+2r_G+3d_Gh_G\right].\ee
Notice that for $z=\pm1$ (so in particular when the theory has $\CN=2$ supersymmetry) the global contribution to the trial a 
central charge has a critical point for $\epsilon=\mp1/3$. It is a minimum for $g=0$ and a maximum for $g>1$. In the following 
we will concentrate on the $A_{N-1}$ theory, so $r_G=N-1$, $h_G=N$ and $d_G=N^2-1$.

\subsection{Determination of the contribution from regular punctures}

Let us consider now a black puncture. Locally the puncture preserves $\CN=2$ supersymmetry and the two $U(1)$ 
symmetries which rotate $x_1$ and $x_2$ should be interpreted as $\frac{1}{2}R_{\CN=2}$ and $I_3$ (the Cartan generator of 
the $SU(2)$ R-symmetry) respectively. We then find $R_0=\frac{1}{2}R_{\CN=2}+I_3$ and $F=\frac{1}{2}R_{\CN=2}-I_3$, which implies 
$R_{\epsilon}=\frac{1+\epsilon}{2}R_{\CN=2}+(1-\epsilon)I_3$. The formula for the trial a central charge (\ref{ac})
 can now be easily written in terms of $R_{\CN=2}$ and $I_3$. Using then the formulas (see \cite{yujidist, TMWG})
\be\label{anomalyN=2}\Tr R_{\CN=2}=\Tr R_{\CN=2}^3= 2n_v - 2n_h;\quad 2\Tr R_{\CN=2}I_3^2= n_v,\ee
we can express the contribution to 
the trial central charge in terms of $n_v$ and $n_h$, the effective number of vectormultiplets and hypermultiplets introduced before:  
\be a_{pb}(\epsilon)=\frac{3}{128}\left[\epsilon^3(12n_v-3n_h)-9\epsilon^2n_h-\epsilon(4n_v+5n_h)+8n_v+n_h\right].\ee
Notice that regardless of the values of $n_h$ and $n_v$ (as long as they are positive), the above expression is always maximized for 
$\epsilon=-1/3$. Using now (\ref{nseff}) we can determine the contribution to $a(\epsilon)$ of any regular puncture. 
For full and simple punctures we have respectively
\be\label{amax}a_{fb}(\epsilon)=\frac{3}{128}[\epsilon^3(6N^3-6N^2)+\epsilon^2(6N-6N^3)-\epsilon(6N^3-2N^2-4N)+6N^3-4N^2-2N],\ee
\be\label{amin}a_{sb}(\epsilon)=\frac{3}{128}[\epsilon^3(9N^2-12)-\epsilon^29N^2-\epsilon(9N^2-4)+9N^2-8].\ee

Clearly, the same argument can be applied for red punctures. The only difference is that 
the roles of $R_{\CN=2}$ and $I_3$ are interchanged with respect to the previous case. This flips the sign of $F$, or equivalently 
the sign of $\epsilon$. The rest of the argument is not modified so, we can conclude that $a_{pr}(\epsilon)=a_{pb}(-\epsilon)$. 
In particular $a_{pr}(\epsilon)$ is maximized for $\epsilon=1/3$.

\section{Checks of the proposal}

In this section we will perform some checks of our proposal. First of all, we will see that we can recover the known constraint for 
$\CN=2$ theories of class $\mathcal{S}$: the SW differential has scaling dimension one. We will then consider theories 
associated with three-punctured spheres and check that our prescription allows us to recover the field theory interpretation proposed 
in \cite{dann1}, \cite{kazuya} following \cite{TMWG}. We will then consider lagrangian theories that can be constructed by 
``connecting'' these basic building blocks. A subclass of these has already been studied in \cite{BB}. We will first recover the 
results of this paper and then we will study in detail two models which have not been considered in \cite{BB}. 
Our analysis is essentially an extension of the argument given in \cite{revt} (section 12.5). 

\subsection{$\CN=2$ theories with regular punctures}

When all the spheres and all the punctures are of the same type, the $CY_3$ geometry becomes $T^*\mathcal{C}\times\mathbb{C}$ and 
we have enhanced $\CN=2$ supersymmetry. There are two possible cases: $p=2g-2$, $q=0$ and correspondingly $\Omega=dx_2\wedge\Omega_2$ 
or $p=0$, $q=2g-2$ and $\Omega=dx_1\wedge\Omega_2$, where $\Omega_2$ is the holomorphic two-form on $T^*\mathcal{C}$. As we explained 
before, the trial a central charge is given by the sum of a global term (extracted from the anomaly polynomial) and local 
contributions from each puncture and as we have noticed before, all these quantities have a critical point at $\epsilon=-1/3$ 
and $\epsilon=1/3$ respectively. Consequently, these are the values at 
which the trial a central charge is maximized. This is obvious when $g\neq0$, since both the global and local terms are maximized for 
those values of $\epsilon$. In the case of the sphere, this is still true provided one adds sufficiently many punctures. 
After all this is not surprising, since a constraint on the number of punctures is known to arise in $\CN=2$ class $\mathcal{S}$ theories 
(see \cite{GMT} for a detailed discussion on this point). 
The precise constraint can be found recalling that the positivity of $n_v$ and $n_h$ is equivalent to the unitarity bounds 
found in \cite{ST}. We thus demand that the theory satisfies this constraint. The global contributions to $n_v$ and $n_h$ from 
the sphere are respectively \cite{CD} 
$$n_v=-\frac{4}{3}N^3+\frac{N}{3}+1=-\sum_{k=2}^{N}(2k-1)^2;\quad n_h=n_v+N-1.$$ We then only need to require the positivity of 
$n_v$. We see from (\ref{nseff}) that the correct requirement is 
\be\label{ineq}\sum_i\sum_{k=2}^{N}(2k-1)p_k^i\geq\sum_{k=2}^{N}(2k-1)^2.\ee
Using the formulas of the previous section it is straightforward to check that the second derivative of the trial a central 
charge is strictly negative when the above inequality is satisfied. Notice that the above bound is less restrictive than the constraint 
$$\sum_ip_k^i\geq2k-1$$ imposed in \cite{CD}, which comes from the requirement that the subspace of dimension k operators of the Coulomb 
branch has positive dimension for every k. There are indeed models that violate the latter bound, such as the higher rank version of 
Minahan-Nemeschansky $E_n$ theories (see \cite{BBT} and references therein).

For these values of 
$\epsilon$ the symmetry $R_{\epsilon}$ becomes the well-known combination $$R_{\epsilon}=\frac{1}{3}R_{\CN=2}+\frac{4}{3}I_3,$$ 
giving the $U(1)$ R-symmetry of the $\CN=1$ subalgebra. Using the formula relating the R-charge to the scaling dimension of 
chiral primary operators $$D(\mathcal{O})=\frac{3}{2}R(\mathcal{O}),$$ we find that in both cases $$D(\Omega_2=d\lambda_{SW})=1,$$ 
which is indeed equivalent to the requirement that the SW differential has scaling dimension one, as expected\footnote{A possible alternative argument, 
which also applies for theories with irregular punctures (and a nontrivial Higgs branch), is the following: as we said in this case 
$\Omega=d\lambda\wedge dx_2$. We also know that the positions of the branes in the transverse $\mathbb{R}^3$ (so in particular 
along $x_2$) describe the motion along the Higgs branch of the theory \cite{DK}. By imposing that the scaling dimensions of $x_2$ 
and the moment map $\mu$ associated with the global symmetry of the theory are the same, as is the case for $\CN=1$ theories (see 
\cite{kazuya}), we immediately reach the desired conclusion: in $\CN=2$ theories the moment map is in the same supermultiplet as 
the conserved current, so its dimension is equal to the canonical one (namely two). Combining then the conditions $[x_2]=2$ and 
$[\Omega]=3$ we find $[\lambda]=1$.}.

\subsection{Three-punctured spheres}

We would like to point out that the 
identification of $\CN=1$ class $\mathcal{S}$ theories with the models studied in \cite{TMWG} is essentially based on duality arguments 
in four dimensions: rather as in the $\CN=2$ case, we expect Seiberg duality and its generalizations (see \cite{TMWG}) to have a geometric 
realization. This is discussed in detail in \cite{dann1} and is indeed true, provided the basic building blocks (three-punctured 
spheres) coincide with the field theories described in \cite{TMWG}. Using the formulas of the previous section it is 
straightforward to compute the trial central charges $a(\epsilon)$ and $c(\epsilon)$ for three-punctured spheres. We will now see 
that these quantities coincide precisely with the expressions one would get for the corresponding field theories studied in 
\cite{TMWG}. This provides a direct six-dimensional check of the identification proposed in \cite{dann1,kazuya}.

Consider a three-punctured sphere (let's say black in our terminology). When the punctures are all full, this describes $T_N$ 
theory plus a chiral multiplet $M$ in the 
adjoint representation of the corresponding $SU(N)$ symmetry for each red puncture (see \cite{dann1}). 

\begin{figure}
\centering{\includegraphics[width=.4\textwidth]{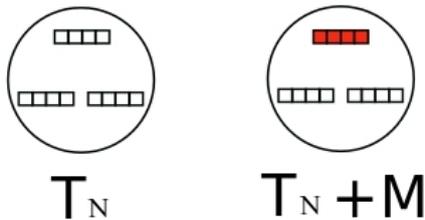}} 
\caption{\label{tienne}\emph{On the left we have the three-punctured sphere representing $T_N$ theory. On the right we turned one 
full puncture of black type into one of red type (we call this process ``rotation''). The resulting sphere describes $T_N$ coupled 
to a chiral multiplet $M$ in the adjoint of $SU(N)$. There is also the superpotential term $\Tr\mu M$.}}
\end{figure}

The trial a central charge for the $T_N$ theory is 
obtained by summing the contribution of three full punctures of black type and the global contribution with $z=1$. The result is 
\be\label{tn}\frac{3}{128}[\epsilon^3(6N^3-18N^2+12)+\epsilon^2(6N-6N^3)-\epsilon(6N^3-6N^2-4N+4)+6N^3-12N^2-2N+8].\ee 
Setting $\epsilon=-1/3$ (which is the value at which (\ref{tn}) is maximized) we recover the a central charge of $T_N$ theory.
If we now ``rotate'' one of the three punctures (see figure \ref{tienne}), we should set $z=0$ 
in (\ref{apoly})  and add the contribution $2a_{fb}+a_{fr}$. 
The difference between this quantity and (\ref{tn}) is equal to $$\frac{3}{32}(N^2-1)(3\epsilon^3-\epsilon),$$ which is precisely the contribution 
of a chiral multiplet in the adjoint of $SU(N)$ with R-charge $1+\epsilon$. Also the variation of the c central charge is compatible 
with this interpretation. 

We now see how the prescription mentioned in the previous section should be modified: in this case the puncture and the sphere are of 
different types and the scaling dimension of the coordinate which does not diverge at the puncture (in this case $x_1$) has to be 
identified with that of the chiral multiplet $M$, not that of the moment map (this was already noticed in \cite{kazuya}). Notice that the superpotential
$\Tr\mu M$, where $\mu$ is the moment map associated with the puncture, is exactly marginal: the three $SU(N)$ moment maps all have 
scaling dimension $3/2-3/2\epsilon$, whereas the multiplet $M$ has dimension $3/2+3/2\epsilon$. This is the superpotential term 
predicted in \cite{TMWG,dann1}. 

If we now rotate another puncture the variation in the 
trial central charge is always the same: $z$ decreases by one and we should add $a_{fr}-a_{fb}$. The maximum of the trial central 
charge is always at $\epsilon=-1/3$.

\begin{figure}
\centering{\includegraphics[width=.18\textwidth]{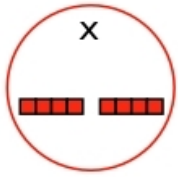}} 
\caption{\label{sfera3}\emph{The three-punctured sphere describing the theory we obtain starting from $T_N$ plus a chiral multiplet 
$M$ and giving to $M$ a nilpotent vev with a Jordan block of dimension $N-1$. The cross again denotes a simple puncture.}}
\end{figure}

What happens if we ``rotate'' instead a generic puncture? Our claim is that we obtain a theory first described in \cite{TMWG}: 
we start from $T_N$ coupled to a multiplet $M$ in the 
adjoint of $SU(N)$ (and the superpotential described before) and we close the full puncture by giving a nilpotent vev to $M$ instead of 
the moment map (the rule for determining the nilpotent vev given the Young diagram associated with the puncture is the same as in the 
$\CN=2$ case). Using our formula we can indeed compute the trial a central charge and confirm this interpretation. 
Let's consider as an example the case of a simple 
puncture and a sphere of red type (see figure \ref{sfera3}). Subtracting the trial central charge for 
$T_N$ found before, we get the expression 
\be\label{sferamin} \frac{3}{128}\left[-8+N(2-4\epsilon-6\epsilon^2)+ 
 N^2(13-7\epsilon-9\epsilon^2+3\epsilon^3)-6N^3(1+\epsilon)(1-\epsilon)^2\right].\ee 
We can reproduce this result from the field theory interpretation mentioned above. This just involves a slight modification of 
the analysis performed in \cite{TMWG}: the structure of the puncture is accounted for by the introduction of the 
superpotential (see \cite{TMWG} and \cite{yujivafa} for the details) \be\label{super}\mathcal{W}=\mu_{1,-1,1}+\sum_{j,k} M_{j,-j,k}\mu_{j,j,k}.\ee 
Under $R_0$ and $F$ the fields $\mu$ have charge 1 whereas $M$ has charge 1 and $-1$ respectively. We also have to take into account 
the $U(1)$ generated by $\rho(\sigma_3)$, where $\rho$ describes the embedding of $SU(2)$ 
in $SU(N)$ defining the puncture. The charge 
of the multiplets entering in the superpotential is given by the second subscript in (\ref{super}). 

The candidate $U(1)_R$ in the IR is given by the following combination of these three $U(1)$ groups: 
$$U(1)_{\epsilon}=R_0+\epsilon F+(\epsilon-1)\rho.$$ In \cite{TMWG} the authors evaluated the 't Hooft anomalies using $U(1)_0$. This is 
because they were interested in the theory obtained coupling two copies of the present model through an $\CN=1$ vectormultiplet. 
$U(1)_0$ is then the correct choice. Here we are considering the three-punctured sphere in isolation, so the correct procedure is 
to keep $U(1)_{\epsilon}$ and then apply a-maximization. The final result is 
$$\Tr U(1)_{\epsilon}^3=\Tr\mathcal{R}^3+\frac{3}{2}(\epsilon-1)^3NI_{p}+c_3;\quad \Tr U(1)_{\epsilon}=\Tr\mathcal{R}+c_1,$$ where 
$\Tr\mathcal{R}$ and $\Tr\mathcal{R}^3$ represent the contributions from $T_N$ theory, $I_{p}$ is the embedding index of 
$SU(2)$ in $SU(N)$ (see e.g. \cite{AS} for the definition of embedding index) and $c_1$ and $c_3$ are the contributions from the multiplets $M_{j,-j,k}$. Their charge under $U(1)_{\epsilon}$ 
is $1-\epsilon+(1-\epsilon)j$, so we immediately find 
$$c_1=N-2-\epsilon(N+1)+\sum_{j=1}^{N-2}[(1-\epsilon)j-\epsilon]=\frac{(N+1)(N-2)}{2}-\epsilon\frac{N(N+1)}{2},$$
$$c_3=\frac{(N-2-\epsilon N)^3}{4}-\epsilon^3+\sum_{j=1}^{N-2}[(1-\epsilon)j-\epsilon]^3.$$
\begin{figure}
\centering{\includegraphics[width=.17\textwidth]{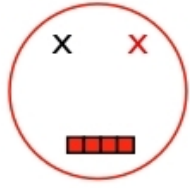}} 
\caption{\label{tiennemod}\emph{Sphere with a full and two simple punctures. Starting from a bifundamental of $SU(N)$ plus 
a chiral multiplet $M$ in the adjoint of one $SU(N)$ group (again with superpotential $\Tr\mu M$), we can obtain this theory 
giving a nilpotent vev to $M.$}}
\end{figure}
If we subtract the contribution from $T_N$ theory we are left with 
$$a(\epsilon)=\frac{3}{32}\left[\frac{9}{2}(\epsilon-1)^3NI_{p}+3c_3-c_1\right],$$ which is precisely equal to (\ref{sferamin}).

We can repeat the exercise for a sphere with two simple punctures and a full one (see figure \ref{tiennemod}). We have to set to zero the parameter $z$ in 
(\ref{apoly}) and add the contributions of two simple punctures (one red and one black) and a full one. From the above discussion, 
we expect the resulting theory to be a 
bifundamental of $SU(N)\times SU(N)$ plus a chiral multiplet $M$ in the adjoint of one $SU(N)$ with the addition of the superpotential 
(\ref{super}). In this case $\mu_{i}^{j}=\widetilde{Q}_i^aQ_a^j$. The charges of $Q$ and $\widetilde{Q}$ under $U(1)_{\epsilon}$ 
are $$R_{\epsilon}(Q_i)=\frac{1+\epsilon}{2}+(\epsilon-1)(\frac{N}{2}-i)\;\; i=1,\dots, N-1;\quad R_{\epsilon}(Q_N)=\frac{1+\epsilon}{2},$$
$$R_{\epsilon}(\widetilde{Q}_i)=\frac{1+\epsilon}{2}+(\epsilon-1)(i-\frac{N}{2})\;\; i=1,\dots, N-1;\quad R_{\epsilon}(\widetilde{Q}_N)=\frac{1+\epsilon}{2}.$$
Combining the contributions from $Q$, $\widetilde{Q}$ and $M$ to the trial central charge we get precisely the expected result, 
thus matching the geometric computation. 

Notice that some of these operators violate the unitarity bound for the value of $\epsilon$ which maximizes the trial a central charge. 
As we mentioned in the introduction, we interpret this as evidence that the ``offending'' operators decouple and become free. This is not surprising after all, since we 
expect free multiplets to arise after the closure of the puncture (see section 2). However, this illustrates an important point: 
the corresponding emergent $U(1)$ symmetries are not manifest from the curve and in order to correctly take this phenomenon into 
account, we should modify ``by hand'' the trial a central charge following the procedure described in \cite{kutasov}.

\subsection{Lagrangian theories}

\subsubsection{Models with full and simple punctures}

It is well known that a hypermultiplet in the bifundamental of $SU(N)\times SU(N)$ has a class $\mathcal{S}$ realization: it is described by 
a sphere with two full punctures and a simple one. Let's consider a collection of $l$ copies of this theory and let's say $p$ of 
them are of black type (see figure \ref{lin} for an example). As we already explained, connecting them corresponds to gauging the 
diagonal combination of the $SU(N)$ symmetries carried by the maximal punctures. When the spheres are of the same type the corresponding 
vectormultiplet is $\CN=2$, otherwise it is $\CN=1$. The resulting theory is described by a sphere with l simple punctures 
and two full ones. The number of gauge groups is $l-1$. This class of theories was studied field theoretically in 
\cite{BB}. We will now check that our procedure allows us to recover that result. 

First of all we need to evaluate the global contribution from the anomaly polynomial. Each sphere gives a +1 contribution to the Chern 
number of the corresponding line bundle and each minimal puncture gives a -1 contribution. These two contributions clearly compensate each 
other and we are left with the contributions from the maximal punctures. If they are both 
black then we get $z=1$, if they are both red we find $z=-1$ and if they are different we get $z=0$. Our parameter z thus plays the role of 
the variable $k$ in \cite{BB}. Using (\ref{amax},\ref{amin}) we can easily add the contribution from the punctures. The final result is 
$$a(\epsilon)=\frac{3}{128}(A_3\epsilon^3+A_2\epsilon^2+A_1\epsilon+A_0),$$ where 
$$A_3=z(12-12N^2)+(2p-l)(9N^2-12);\quad A_2=-9N^2l,$$
$$A_1=z(4N^2-4)-(2p-l)(9N^2-4);\quad A_0=8-8N^2+l(9N^2-8).$$
The terms proportional to $z$ are obtained combining the contributions from the anomaly polynomial and maximal punctures. The rest 
comes from minimal punctures. Punctures with different signs contribute the same amount to $A_2$ and $A_0$, so the final expression is 
just l times the corresponding term in (\ref{amin}). The contributions to $A_3$ and $A_1$ have instead opposite sign, so the final 
result is clearly given by the term appearing in (\ref{amin}) times $p-(l-p)$. 
This fits perfectly with the field theory analysis of \cite{BB} (formula (3.10))\footnote{Notice that the sign convention in \cite{BB} 
differs from ours. Their $k$ is our $-z$ and $zl=p-q$ in that paper corresponds to $-(2p-l)$ in our notation.}. 

\subsubsection{Linear quivers with more general regular punctures}

As is well known, $\CN=2$ linear quivers of unitary groups are part of the so called class $\mathcal{S}$ theories, and in this 
language they are associated with a sphere with an arbitrary number of minimal punctures and two additional generic punctures. The 
lengths of the rows of the Young diagrams of the generic punctures encodes the ranks of the various gauge groups and the requirement 
of scale invariance fixes in turn the matter content of the theory. The same result can be derived directly in field theory exploiting 
the fact that a generic puncture can be obtained starting from a maximal puncture and giving a nilpotent vev to the corresponding moment 
map (which is simply the meson field for lagrangian theories). Expanding the superpotential around the new vacuum (as we did  
before when we discussed three punctured spheres) we get several quadratic terms and integrating out the corresponding 
massive fields we recover the expected linear quiver \cite{revt}. 

In this section we would like to understand the generalization of this story to $\CN=1$: what kind of $\CN=1$ model do we get if we 
replace the maximal punctures of the previous section with more general ones? The answer can be determined by giving the 
proper vev to the matter fields as in the $\CN=2$ case\footnote{We thank K. Maruyoshi for discussions about this point.}. 
The main difference is that the superpotential in this case is quartic instead 
of cubic and by expanding around the vev we will also get cubic couplings among the various fields. There is an obvious test for 
our result: we can match the anomalies with those derived from the geometric setup.

\begin{figure}
\centering{\includegraphics[width=.8\textwidth]{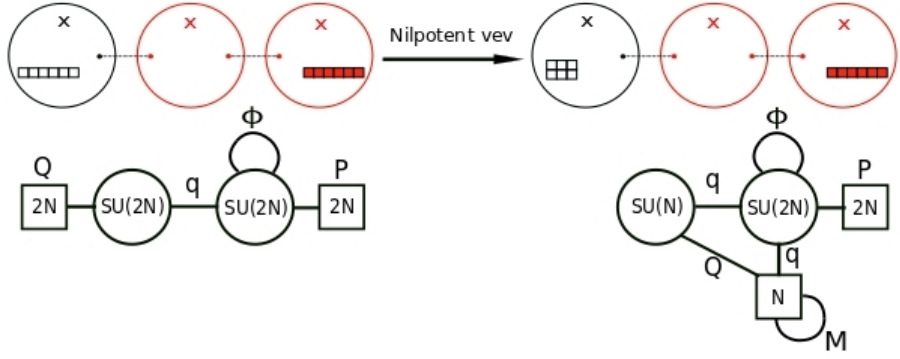}} 
\caption{\label{quivern}\emph{On the left we have the linear quiver of $SU(2N)$ gauge groups (in this case $N=3$). One gauge group is 
$\CN=1$ and the other $\CN=2$. The boxes denote the chiral multiplets in the fundamental and the line between gauge groups the 
bifundamental multiplets. We denote the fundamentals as $Q$ and $P$ and the bifundamental as $q$ (indeed we also have $\widetilde{Q}$, $\widetilde{P}$ and 
$\widetilde{q}$ multiplets). Giving the nilpotent vev we get the theory on the right. Each matter field is denoted with the same 
letter as the parent matter field on the left. We denote as $M$ the chiral multiplet in the adjoint of $SU(N)$.}}
\end{figure}

We will focus on a specific example: the $A_{2N-1}$ theory compactified on a sphere with three minimal punctures, one maximal and 
one labelled by the partition $(2^N)$ (see figure \ref{quivern}). The contribution to $n_v$ and $n_h$ from this puncture are 
$$n_v=\frac{16}{3}N^3-5N^2-\frac{N}{3};\quad n_h=\frac{16}{3}N^3-4N^2-\frac{4}{3}N.$$ 

Starting from the theory with two maximal punctures, we give the nilpotent vev $$\widetilde{Q}Q=I_N\otimes\left(\begin{array}{cc}
  0 & 1 \\
  
  0 & 0
 \end{array}\right),$$ where $I_N$ is the $N\times N$ identity matrix. Modulo flavor and gauge rotations we can bring 
 $Q$ and $\widetilde{Q}$ to the form
\be Q=\left(\begin{array}{c|c}
  0 & I \\
  \hline
  0 & 0
 \end{array}\right);\quad \widetilde{Q}=\left(\begin{array}{c|c}
  I & 0 \\
  \hline
  0 & 0
 \end{array}\right).
\ee In the above equations we decomposed $Q$ and $\widetilde{Q}$ matrices into $N\times N$ blocks. In the following they will be denoted 
as $Q_{ij}$ and $\widetilde{Q}_{ij}$ ($i,j=1,2$). The equations of motion are solved by setting to zero the vev of the other fields. 

Clearly these vevs break the $SU(2N)$ gauge and flavor symmetries to $SU(N)$, with the global $SU(N)$ realized as a diagonal combination 
of gauge and flavor transformations\footnote{Our convention is that $Q$ ($\widetilde{Q}$) transforms in the fundamental (antifundamental) 
of the gauge symmetry and in the antifundamental (fundamental) of the flavor symmetry.}. As a result, the $q$ and $\widetilde{q}$ 
fields now transform non trivially under the global $SU(N)$ symmetry. The resulting gauge and flavor quantum numbers can be summarized 
as follows: 
\begin{center}\begin{tabular}{|c|c|c|}
\hline
 & \bf{fundamental} & $Q_{21}$, $Q_{22}$, $q_{21}$, $q_{22}$ \\
Gauge $SU(N)$ & & \\
 & \bf{antifund.} & $\widetilde{Q}_{12}$, $\widetilde{Q}_{22}$, $\widetilde{q}_{12}$, $\widetilde{q}_{22}$ \\
\hline 
 & \bf{fundamental} & $\widetilde{Q}_{12}$, $\widetilde{Q}_{22}$, $q_{11}$, $q_{12}$ \\
Flavor $SU(N)$ & \bf{antifund.} & $Q_{21}$, $Q_{22}$, $\widetilde{q}_{11}$, $\widetilde{q}_{21}$ \\
 & \bf{adj.} & $Q_{11}$, $Q_{12}$, $\widetilde{Q}_{11}$, $\widetilde{Q}_{21}$ \\
\hline
\end{tabular}
\end{center}
The fields not appearing in the table are just uncharged.

Expanding the quartic superpotential $\Tr (Q\widetilde{Q})_0(q\widetilde{q})_0$ (where $()_0$ indicates the traceless part) around 
this vacuum we find 
\be\begin{split}\label{supp}\mathcal{W}=& \Tr[\sqrt{2}M(q\widetilde{q})_{11}+\widetilde{Q}_{22}(q\widetilde{q})_{21}+ 
Q_{21}(q\widetilde{q})_{12}]\\
& -\frac{1}{2N}\sqrt{2}\Tr M(\Tr(q\widetilde{q})_{11}+\Tr(q\widetilde{q})_{22})+\text{quartic},\end{split}\ee 
where we redefined $M=(Q_{11}+\widetilde{Q}_{21})/\sqrt{2}$ and $\widetilde{M}=(Q_{11}-\widetilde{Q}_{21})/\sqrt{2}$. In this particular 
case the superpotential is cubic because the vev of $Q\widetilde{Q}$ is zero. In more general cases we will also get quadratic 
terms which give a mass to some matter fields. Clearly 
this superpotential breaks the R-symmetry of the original theory, which mixes with the $U(1)$ generated by $\rho(\sigma_3)$
to give the infrared R-symmetry. We should now require the cubic terms in $\mathcal{W}$ to have R-charge two. This singles out 
the combination 
$$R_{\text{new}}=R_{\epsilon}+\frac{\epsilon-1}{2}\rho(\sigma_3).$$ With this assignment all the $Q$ and $\widetilde{Q}$ fields 
appearing in the superpotential have R-charge $1-\epsilon$ and the others have zero R-charge. The remaining matter fields are 
unaffected. The chiral fields with zero R-charge apparently violate the unitarity bound. As explained in \cite{Seiberg} this means that they 
become free and decouple. The same occurs to the field $\widetilde{M}$, since it is not charged under the gauge group and 
does not couple to any other field. 

The a and c central charges of this model can be easily evaluated just summing the contributions from the various matter fields: 
$M$, $Q$, $\widetilde{Q}$ and $\Phi$ (the multiplet in the adjoint of $SU(2N)$) have R-charge $1-\epsilon$. $q$, $\widetilde{q}$, 
$P$ and $\widetilde{P}$ have charge $(1+\epsilon)/2$. Including the gauginos whose R-charge is indeed one we get 
$$\Tr R_{\text{new}}^3=\epsilon^3(1-5N^2)-6N^2\epsilon^2+6N^2\epsilon+3N^2-2,$$
$$\Tr R_{\text{new}}=\epsilon(N^2+1)-3N^2-2.$$ The trial a central is maximized at $\epsilon=1/3$. 

These anomalies can also be computed using the formulas of section 3: summing the local contribution from the various punctures 
with the global 
\begin{figure}
\centering{\includegraphics[width=.8\textwidth]{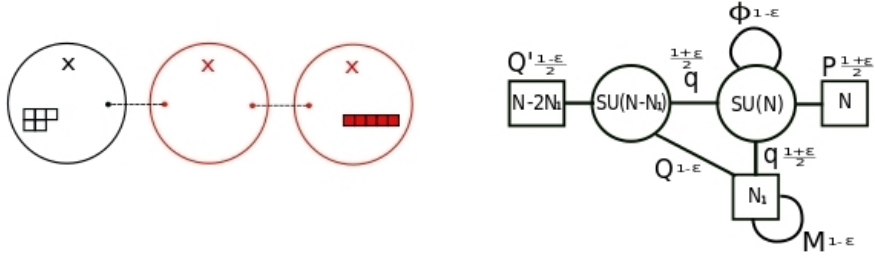}} 
\caption{\label{quivern1}\emph{On the left we have the geometric picture of the gauge theory on the right. In this case $N=5$ and 
$N_1=2$. The $SU(N-N_1)$ and $SU(N)$ vectormultiplets are $\CN=1$ and $\CN=2$ respectively. Next to each matter field we have 
indicated its R-charge.}}
\end{figure}
contribution (with $z=0$) we get exactly the same result, thus confirming our interpretation. Notice that, since 
the coefficients of the various powers of $\epsilon$ match, the triangle anomalies involving the $U(1)$ groups
$R_0-\frac{1}{2}\rho(\sigma_3)$ and $F+\frac{1}{2}\rho(\sigma_3)$ match as well. 

This analysis can be easily generalized to the case of a puncture labelled by the partition $(2^{N_1},1^{N-2N_1})$ (with $N_1<N/2$). This reduces to the 
previous case when $N$ is even and $2N_1=N$. We get a model whose matter content is represented in figure \ref{quivern1}, with a superpotential 
generalizing (\ref{supp}). The vev of the $Q$, $\widetilde{Q}$ matter fields can be put in the form 
\be Q=\left(\begin{array}{c|c|c}
  0 & I & 0\\
  \hline
  0 & 0 & 0\\
  \hline 
  0 & 0 & 0
 \end{array}\right);\quad \widetilde{Q}=\left(\begin{array}{c|c|c}
  I & 0 & 0 \\
  \hline
  0 & 0 & 0\\
   \hline
  0 & 0 & 0
 \end{array}\right),\ee 
where the first two blocks have dimension $N_1$. The infrared R-symmetry is again 
$$R_{\text{new}}=R_{\epsilon}+\frac{\epsilon-1}{2}\rho(\sigma_3).$$ 
The $R_{\text{new}}$ anomalies, computed either from the field theory or geometric data, are 
$$4\Tr R_{\text{new}}^3=\epsilon^3(2N_1^2+4-5NN_1-3N^2)+\epsilon^2(9NN_1-9N^2-6N_1^2)+$$ $$\epsilon(3N^2+9NN_1-6N_1^2)+5N^2-5NN_1+2N_1^2-8,$$
$$\Tr R_{\text{new}}=\epsilon(NN_1-N_1^2+1)+NN_1-N^2-N_1^2-2.$$ 
This model has an $SU(N_1)\times SU(N-2N_1)$ flavor symmetry, where both factors are associated with the puncture labelled by the partition 
$(2^{N_1},1^{N-2N_1})$. Using (\ref{ciao}) we can compute the flavor central charges from the above field theory data: 
$$K_{SU(N-2N_1)}=\frac{3}{2}(1+\epsilon)(N-N_1);\quad K_{SU(N_1)}=3\epsilon(N_1+N-N_1)+\frac{3}{2}(1-\epsilon)N=\frac{3}{2}(1+\epsilon)N.$$
Since the puncture is of black type, from the discussion in section 3.3  we find 
$$K_G\delta^{ab}=-3\Tr RT^aT^b=-\frac{3}{2}(1+\epsilon)\Tr R_{\CN=2}T^aT^b.$$ Combining now (\ref{ciao1}) and (\ref{globcc}) we 
find for the puncture of interest 
$$K_{SU(N-2N_1)}=\frac{3}{4}(1+\epsilon)(2N-2N_1)=\frac{3}{2}(1+\epsilon)(N-N_1);\quad K_{SU(N_1)}=\frac{3}{2}(1+\epsilon)N.$$ 
We thus find again agreement between the geometric and field theory pictures. More general linear quivers can be analyzed along the 
same lines (see \cite{BK}). 

These two models represent the simplest nontrivial examples of the object called fan in \cite{BK}. 
The $U(1)$ symmetries acting as rotations on the fiber of the two line bundles, under which the coordinates $x_1$ and $x_2$ 
introduced in section 3 have charge one, correspond respectively to $J_-/2$ and $J_+/2$ of \cite{BK}. With this identification, one 
can easily check that the spectrum and charge assignments given above perfectly agree with those found in the above-mentioned paper. 

Notice also that using the formulas of section 3 we recover the prescription given in \cite{BK} (section 5) for determining the 
$J_{\pm}$ 't Hooft anomalies for generic $\CN=1$ class $\mathcal{S}$ theories. Our analysis can be seen as a direct derivation from six dimensions of this prescriptions.

\section{$\CN=2$ theories deformed by a superpotential} 

In this section we will study $\CN=1$ theories obtained deforming $\CN=2$ models with an $\CN=1$ preserving 
superpotential studied in \cite{noi}. 

\subsection{Mass deformation of $\CN=2$ theories}

In \cite{noi} a prescrition for determining the $\CN=1$ spectral curve for (a class of) $\CN=2$ theories in class $\mathcal{S}$ 
deformed by a superpotential for the adjoint chiral multiplets was given, extending the Type IIA brane construction of \cite{Hori} (see 
also \cite{BTW} for earlier studies on $\CN=1$ deformations of class $\mathcal{S}$ theories). We start from the Seiberg-Witten 
curve for the underlying $\CN=2$ model $F(t,v)=0$, where $t$ is the coordinate on the Gaiotto curve $\mathcal{C}$ and $v$ parametrizes 
the cotangent bundle. With the conventions of \cite{noi}, which we adopt, the SW differential is $\lambda=(v/t)dt$. The 
equation defining the projection of the $\CN=1$ curve on the $(w,t)$-plane, where $w$ parametrizes the trivial bundle, can be 
derived from the $\CN=2$ curve by imposing the boundary conditions $w\sim\mathcal{W}'(v)$ at the punctures, where $\mathcal{W}$ is 
the $\CN=2$-breaking superpotential.

When the superpotential includes only quadratic terms we should impose the boundary condition at the punctures $w\sim v$. The 
proportionality coefficient encodes the mass parameter in the superpotential. Assuming the mass parameters are generic (i.e. there 
are no emergent $U(1)$ symmetries in the infrared) we should impose that the scaling dimensions of $v$ and $w$ are the same, as a 
consequence of the above boundary condition. Since the holomorphic three-form with the present parametrization is 
$$\Omega=dv\wedge dw\wedge\frac{dt}{t},$$ 
we find that both $v$ and $w$ should have dimension 3/2, or equivalently R-charge one. This corresponds to setting $\epsilon$ 
to zero in (\ref{dimension}). The infrared R-symmetry is then $$R=\frac{1}{2}R_{0}+I_3,$$ in agreement with the general argument 
given in \cite{BT}. This in particular assigns R-charge one to all the moment maps.
Indeed, by integrating out the massive adjoint fields we generate quartic couplings of the form $\Tr\mu_i^2$ 
where $\mu_i$ are the moment maps associated with the gauge symmetries. For generic choices of mass parameters the above R-charge 
assignment is the only one such that none of the above superpotential terms is relevant.

\subsubsection{IR fixed point of SQCD}

Starting from the curve of mass deformed $\CN=2$ SQCD and taking a scaling limit one can extract the curve describing $\CN=1$ 
SQCD \cite{Hori}: let's consider the $\CN=2$ curve in the form (we consider the case $N_f>N$ only) 
$$t^2+tP_N(v)+\Lambda^{2N-N_f}v^{N_f}=0;\quad \lambda=v\frac{dt}{t}$$ 
we can impose the boundary condition 
\begin{itemize}
 \item $w\rightarrow0;\quad v^{N_f-N}\sim t$ for $t\rightarrow0$;
 \item $v\rightarrow0;\quad w^N\sim t$ for $t\rightarrow\infty$.
\end{itemize}
Considering these boundary conditions and after the rescaling $t\rightarrow t/\mu^N$ (where $\mu$ is the mass of the chiral 
multiplet in the adjoint) we find the curve 
\begin{equation}\begin{cases} v^{N_f-N}\Lambda_{\CN=1}^{3N-N_f}=t\\
 w^N=t\\
 vw=0
\end{cases}\quad \Omega=dv\wedge dw\wedge\frac{dt}{t}. \end{equation}
The third equation simply tells us that the curve develops two branches. From these expressions for the curve and differential 
we immediately find $$[v]=3\frac{N}{N_f};\quad [w]=3\frac{N_f-N}{N_f}.$$ A useful observation at this stage is that the scaling 
dimension of $w$ is the same as the scaling dimension of the meson $\widetilde{Q}Q$: as was done in \cite{Hori}, one can verify 
that in the massive theory the limit of $w$ for small $t$ is proportional to the vev of the meson (this also fits well with the 
discussion in \cite{kazuya}). We thus recover the 
well-known result \cite{Seiberg} $$[\widetilde{Q}Q]=3\frac{N_f-N}{N_f}.$$

\subsection{$\CN=1$ deformation of Argyres-Douglas theories}

Let's first of all explain what we mean by $\CN=1$ deformation: we can see the Argyres-Douglas theory (we will consider only the $A_{N-1}$ 
case) as a singular point in the Coulomb branch of $SU(N)$ SYM theory (that's how it was originally discovered in \cite{AD}). 
The SW curve describing the theory has the form $y^2=x^N+\sum_{k=2}^{N}x^{N-k}u_k$. The parameters $u_k$ with $k>N/2+1$ 
represent chiral operators in the SCFT which are inherited from the $\Tr\Phi^k$ operators of the gauge theory. The other 
parameters are interpreted as the corresponding coupling constants. 

Our strategy is to start from SYM theory and turn on a polynomial superpotential for the chiral multiplet in the adjoint 
$$\mathcal{W}=\int d^2\theta\Tr\Phi^k.$$ It is known that when $k>N/2+1$ this perturbation does not lift the AD point\footnote{This 
can be seen for example by analyzing the factorization condition for the Dijkgraaf-Vafa curve. See e.g. \cite{ES} for details. When $N$ is 
even also $k=N/2+1$ is allowed. We will not treat this case.}. 
We interpret this operation as turning on the deformation 
$$\mathcal{W}=\int d^2\theta u_k$$ 
in the AD SCFT. This is indeed the most natural guess and we will momentarily give evidence for it. For the moment just notice that the 
bound on k for the exponent in the superpotential is the same we get in the $\CN=2$ theory (only when $k>N/2+1$ $u_k$ corresponds 
to a chiral operator in the CFT).

The curve for the case $k=N$ was given explicitly in \cite{noi}. This can be generalized to
\begin{equation}\begin{cases} x^N=\Lambda^N(t+1)^2/t\\
 w^N=\Lambda^{Nk-N}t^{N+1-k}(t+1)^{2k-2-N}\\
 x^{N+1-k}w=\Lambda^N(t+1)
\end{cases}\quad \Omega=dx\wedge dw\wedge\frac{dt}{t}. \end{equation}
This formula reproduces the correct Dijkgraaf-Vafa curve (see \cite{FDSW} and references therein) $$w^2=x^{k-1}w-\Lambda^Nx^{2k-2-N}$$ and the correct asymptotics 
$w\sim v^{k-1}$ for large $t$. Expanding around the singular point we get the curve 
\begin{equation}\begin{cases} x^N=z^2\\
 w^N=z^{2k-2-N}\\
 x^{N+1-k}w=z
\end{cases}\quad \Omega=dx\wedge dw\wedge dz. \end{equation}
Imposing now the constraint $[\Omega]=3$ we find 
$$[x]=\frac{3}{k};\quad [w]=\frac{6k-6-3N}{2k};\quad [z]=\frac{3N}{2k}.$$ 
The fact that we can consistently assign scaling dimensions to the coordinates suggests that these $\CN=1$ deformations lead to 
nontrivial IR fixed points. From now on we will assume this is the case and call the resulting SCFTs $I_{N,k}$.

In conclusion, we find that $u_k$ has scaling dimension three confirming our expectation: at the IR fixed point the operator
$$\int d^2\theta u_k$$ which initiated the flow from the AD point becomes exactly marginal. The operators $u_n$ (for $n>N/2+1$) 
become operators of the new $\CN=1$ theory and their scaling dimension is $3n/k$ (of course those with $n>k$ are irrelevant). 

Under the assumption that there are no emergent $U(1)$ symmetries, the R-symmetry in the infrared is a combination of $R_{\CN=2}$ and $I_3$ of the underlying $\CN=2$ theory. From our assignment of scaling 
dimensions we can immediately determine it: at the AD point $x$ has charge $4/(N+2)$ under $R_{\CN=2}$ and zero under $I_3$, whereas 
$w$ has charge one under $I_3$ and is uncharged under $R_{\CN=2}$ (this is because $R_{\CN=2}$ is to be identified with 
the group of rotations in the $x$-plane, so does not act on $w$). Combining these charge assignments we find 
\be\label{car} R_{IR}=\frac{N+2}{2k}R_{\CN=2}+\frac{2k-2-N}{k}I_3,\ee 
which is precisely the combination preserved by the superpotential term $$\mathcal{W}=\int d^2\theta u_k.$$ 

Exploiting this fact we can now compute the a and c central charges for the $\CN=1$ deformed AD theory. Using the relations \cite{ST, BT} 
\be\label{scft}\Tr R_{\CN=2}^3=\Tr R_{\CN=2}=48(a-c);\quad \Tr R_{\CN=2}I_3^2=4a-2c,\ee where a and c are the central charges of the AD 
theory, we can immediately evaluate $$a_{IR}=\frac{3}{32}(3\Tr R_{IR}^3-\Tr R_{IR});\quad c_{IR}=\frac{1}{32}(9\Tr R_{IR}^3-5\Tr R_{IR}).$$ 

Using (\ref{car}) and (\ref{scft}) we find 
$$a_{IR}=\frac{3}{32}\frac{N+2}{8k^3}\left[48(a-c)(3(N+2)^2-4k^2)+72(2a-c)(2k-2-N)^2\right],$$ 
$$c_{IR}=a_{IR}-\frac{N+2}{2k}3(a-c).$$
The a and c central charges for the AD theory were found in \cite{ST}. Exploiting this result we get 
\begin{itemize}
 \item For $N=2r+1$
\end{itemize}
\be\label{apar} a_{IR}=\frac{3}{32}\frac{r[3(2+3r)(3+2r)^2-9k(3+2r)(3+4r)+4k^2(7+9r)]}{k^3}.\ee 
\begin{itemize}
 \item For $N=2r+2$
\end{itemize}
\be\label{adis} a_{IR}=\frac{3}{16}\frac{2k^2+rk^2-3(2+r)^3+9r(3+2r)(k-2-r)^2}{k^3}.\ee

We can now perform a nontrivial consistency check exploiting the fact that the theories $I_{N,k}$ labelled by the same $N$ are all related 
by an RG flow: if $k_1>k_2$ we can start from the AD point and turn on the deformation $\int d^2\theta u_{k_1}$ thus flowing to the 
theory $I_{N,k_1}$. If we now turn on the deformation $\int d^2\theta u_{k_2}$ the flow initiates again and we reach the theory 
$I_{N,k_2}$. The a-theorem \cite{KS} then tells us that the a central charge should decrease along the flow and this in turn is equivalent 
to the requirement that $a_{IR}$ is non decreasing as a function of k. One can check that in the allowed range for k ($N/2+1<k<N$) 
the derivative with respect to $k$ of the above two functions is positive for any value of r. Notice that the ratio $a_{IR}/a$ for 
the theory $I_{N,N}$ (where a is again the central charge of the $\CN=2$ theory) tends to $27/32$ in the large N-limit, as in the 
case of quadratic superpotential studied in \cite{BT}. Indeed, $a-a_{IR}$ is positive for every $N$ and $k$ as it should be (just because 
we have an RG flow from the $\CN=2$ theory to $I_{N,k}$).

This is not the end of the story though: as we mentioned earlier, the AD theory admits the relevant deformations 
$$\int d^2\theta d^2\tilde{\theta}u_{N-j+2}u_j,\quad (j>N/2+1)$$ where $d^2\theta d^2\tilde{\theta}$ denotes the integral over half of the 
$\CN=2$ superspace. The key point for us is that after the $\CN=1$ breaking, besides the operators $u_j$, we also have the 
chiral operators $$v_j=\int d^2\tilde{\theta}u_j.$$ 
We can evaluate their scaling dimension exploiting the fact that $\tilde{\theta}$ has charge 1 under 
$R_{\CN=2}$ and $-1/2$ under $I_3$. From (\ref{car}) we then find 
$$R_{IR}(v_j)=\frac{2k+2j-4-2N}{k}\Longrightarrow[v_j]=\frac{6k+6j-12-6N}{2k}.$$ Not surprisingly, the coordinates $u_{N-j+2}$ are 
the corresponding coupling constants (notice that $[v_j]+[u_{N-j+2}]=3$). 

For $N$ even it is important to notice that the theory has a Higgs branch and an associated $U(1)$ global symmetry (when $N=4$ 
this enhances to $SU(2)$). The parameter $u_{N/2+1}$, whose dimension is one in the $\CN=2$ theory, is 
the mass parameter associated with this global symmetry. Consequently, in this case we should also include the moment map $\mu$ associated 
to the global symmetry in the list of chiral operators of the theory. This of course is not visible in the $\CN=2$ curve because 
its vev parametrizes the Higgs branch rather than the Coulomb branch of the theory. Its charge under $R_{IR}$ can be easily fixed: 
extended supersymmetry implies that its charges under $R_{\CN=2}$ and $I_3$ are zero and one respectively, so from (\ref{car}) 
we immediately find $$R_{IR}(\mu)=\frac{2k-2-N}{k}\Longrightarrow [\mu]=\frac{6k-6-3N}{2k}.$$ 

We now see that only when 
\be\label{ops} k\geq\frac{3}{4}N+\frac{3}{2}\;\;\text{(for $N$ even)};\quad\quad k\geq\frac{3}{4}N+\frac{3}{4}\;\;\text{(for 
$N$ odd)},\ee all the $v_j$'s and $\mu$ (when it exists) have scaling dimension larger than one. Consequently, for smaller values of k some operators violate 
the unitarity bound. Again, following \cite{Seiberg}, we interpret this as evidence that the operators become free and decouple. Whenever this 
happens, the computation of the a central charge should be modified accordingly: for each operator that violates the unitarity 
bound, we should subtract from (\ref{apar}), (\ref{adis}) the contribution of a chiral multiplet with the same R-charge and add the contribution 
of a free chiral multiplet, whose R-charge is $2/3$. We thus learn that  (\ref{apar}) and (\ref{adis}) are reliable only when 
(\ref{ops}) is satisfied. In particular we find that our assumption regarding the absence of emergent $U(1)$ symmetries is correct only in the 
above range for k. This modification increases the value of the a central charge and is crucial 
for the consistency of our analysis, since (\ref{apar}) and (\ref{adis}) give a negative a central charge for large enough 
$r$ and small $k$.

When $k$ is in the range (\ref{ops}), (\ref{apar}) and (\ref{adis}) pass another consistency check: the a and c central charges 
satisfy the Maldacena-Hofman bound \cite{MH}: 
\be\label{mhb}\frac{1}{2}\leq\frac{a}{c}\leq\frac{3}{2}.\ee
One way to see it is to introduce the effective number of vector and chiral multiplets $n_v$ and $n_{\chi}$: 
\be\label{neff}a=\frac{9n_v+n_{\chi}}{48};\quad c=\frac{3n_v+n_{\chi}}{24}.\ee 
These agree with the number of vector and chiral multiplets for lagrangian theories and their positivity is equivalent to (\ref{mhb}). 
From (\ref{car}) we have $$\Tr R_{IR}=\frac{N+2}{2k}\Tr R_{\CN=2},$$ and the rhs is always negative for Argyres-Douglas theories \cite{ST}. 
Using (\ref{ac}) and the above equation we find 
$$\frac{n_{\chi}}{12}=\frac{9}{32}(\Tr R_{IR}^3-\Tr R_{IR})>a_{IR}>\frac{1}{32}(9\Tr R_{IR}^3-\Tr R_{IR})=\frac{n_v}{4}.$$  
The result then follows from the positivity of the rightmost term, which can be easily proven in the range (\ref{ops}) using 
(\ref{scft}) and the formula for the central charges of Argyres-Douglas theories.

There is one exception to the rule discussed above: when $N=4$ (so the only allowed value 
for $k$ is four) the only operator which violates the unitarity bound is the moment map of the $SU(2)$ flavor symmetry, which transforms in the 
adjoint and consequently describes three chiral multiplets. In this case what happens is that the three chiral multiplets decouple 
simultaneously. From (\ref{apar}), we can see that the a and c central charges for $I_{4,4}$ match those of a theory describing 
three chiral multiplets with R-charge $1/2$. Our procedure thus leads to the conclusion that $I_{4,4}$ is a free theory: it just 
describes three non interacting chiral multiplets. Similarly, we can predict that $I_{3,3}$ and $I_{5,4}$ are free theories 
describing a single chiral multiplet. 

We would like to point out that our conclusions about $\CN=1$ Argyres-Douglas models differ from those of \cite{ES}. Although 
the constraints on the scaling dimensions coming from the curve alone (without taking into account $\Omega$) are the same, we
differ in the interpretation of the $\CN=2$-breaking superpotential: to impose marginality of the superpotential in the IR, the 
authors of \cite{ES} require $[\Tr\Phi^k]=3$, where the dimension of $\Tr\Phi^k$ is extracted from a certain scaling limit of the generalized 
anomaly equations. This procedure leads to the following infrared R-symmetry: 
$$R_{IR}=\frac{N+2}{N+2k}R_{\CN=2}+\frac{4k-4-2N}{N+2k}I_3.$$ 
This combination is not compatible with the RG flow induced by terms of the form $\int d^2\theta u_j$. We believe our analysis is more natural, 
since it leads to a clear field-theoretic interpretation of the $\CN=1$ breaking.

\section{Final remarks}

In this paper we proposed a method to determine the scaling dimension of chiral operators in $\CN=1$ superconformal theories 
obtained compactifying the 6d $\CN=(2,0)$ theory on a Riemann surface with punctures. This can be used to explore the 
properties of nonlagrangian $\CN=1$ superconformal theories such as $\CN=1$ Argyres-Douglas theories in this paper. As in the $\CN=2$ case, 
the fact that the curve allows us to consistently assign scaling dimensions to the operators is an indication that the theory 
in question is superconformal. In general this is not possible, as in the case of SYM theory discussed in section 3. In fact in 
this case the theory is known to become massive in the infrared.

As we have seen, our procedure can be 
considered a generalization of the known prescription for $\CN=2$ theories. When the R-symmetry of the $\CN=1$ theory associated with the 
punctured sphere is not uniquely specified by the symmetries we need to use a-maximization to determine it. In this case our procedure 
allows us to fix the charges of fields under the trial R-symmetry. In the present note we have found some examples of this phenomenon: 
there are two $U(1)$ symmetries manifest from the curve which act as phase rotations on the fiber of the two line bundles. We stress 
that this procedure leads to the correct answer under the assumption that there are no emergent $U(1)$ symmetries which are 
not manifest from the curve (which is the same limitation underlying a-maximization).
We then need to compute the trial central charge and maximize it to find the combination which realizes the exact R-symmetry. 

Once this is done, we need to check whether some chiral fields violate the unitarity bound, which in turn may signal the presence of emergent 
$U(1)$ symmetries the R-symmetry mixes with. These are not manifest from the curve and we should adjust ``by hand'' the trial 
central charge following the procedure described in \cite{kutasov}. We have seen an occurrence of this phenomenon in section 4 
and also in section 5. 

It would be interesting to extend our analysis to theories associated with surfaces with more general types of punctures, which 
do not ``locally preserve'' $\CN=2$ supersymmetry and explore models coming from 6d theories of D or E type.  

\section*{Acknowledgments} 
It is a pleasure for the author to thank Yuji Tachikawa for carefully reading the manuscript and for his helpful comments. 
We also thank Kazunobu Maruyoshi for useful discussions.
This work was partially supported by the ERC Advanced Grant "SyDuGraM", by FNRS-Belgium (convention FRFC PDR T.1025.14 and 
convention IISN 4.4514.08) and by the ``Communaut\'e Fran\c{c}aise de Belgique" through the ARC program.

\bibliographystyle{ytphys}

\end{document}